\documentstyle[12pt,epsfig]{article}

\def\ap#1#2#3{{Ann. Phys. (N.Y.) {\bf #1}, #2 (#3)}}
\def\ijm#1#2#3{{Int. J. Mod. Phys. {\bf A~#1}, #2 (#3)}}
\def\nc#1#2#3{{Nuovo Cimento {\bf #1~A}, #2 (#3)}}
\def\npb#1#2#3{{Nucl. Phys. {\bf B~#1}, #2 (#3)}}
\def\plb#1#2#3{{Phys. Lett. {\bf B~#1}, #2 (#3)}}
\def\prd#1#2#3{{Phys. Rev. {\bf D~#1}, #2 (#3)}}

\def\ptp#1#2#3{{Prog. Theor. Phys. {\bf #1}, #2 (#3)}}

\begin{document}

\sloppy

\centerline{\Large {\bf Chiral predictions for 
\mbox{\boldmath $K_L \rightarrow
\pi^0 \gamma \mu^+ \mu^-$}}}

\medskip
\bigskip\bigskip\medskip

\centerline{\large{John F. Donoghue}}
\medskip

\centerline{\large{Department of Physics and Astronomy}}
\smallskip

\centerline{\large{University of Massachusetts, Amherst, MA~01003, USA}}
\bigskip

\centerline{\large and}
\bigskip

\centerline{\large{Fabrizio Gabbiani}}
\medskip

\centerline{\large{Department of Physics}}
\smallskip

\centerline{\large{Duke University, Durham, NC~27708, USA}}

\vskip 3.0 truecm

\begin{abstract}

We have previously analysed the chiral predictions for
the decay $K_L \rightarrow \pi^0 \gamma e^+ e^-$ and discussed how 
experimental measurements of this process can help our understanding
of chiral theories and of CP tests. Motivated by the possibility that
the analogous muonic process may soon be measurable, we extend this 
calculation to the
decay $K_L \rightarrow \pi^0 \gamma \mu^+ \mu^-$. Branching ratios and
differential branching ratios are calculated and presented. Measurements of
these will help shed some light on the puzzles in our 
understanding of the chiral loop effects which were raised by the rate for 
$K_L \rightarrow \pi^0 \gamma \gamma$.

\end{abstract}
{\vfill UMHEP-451, DUKE-TH-98-160}
\eject

\section{Introduction}

The radiative rare kaon decays
\begin{eqnarray}
K_L & \rightarrow & \pi^0 \gamma \gamma, \nonumber \\
K_L & \rightarrow & \pi^0 \gamma e^+ e^-, \nonumber \\
K_L & \rightarrow & \pi^0 e^+ e^-, \nonumber \\
K_L & \rightarrow & \pi^0 \gamma \mu^+ \mu^-, \nonumber \\
K_L & \rightarrow & \pi^0 \mu^+ \mu^-
\end{eqnarray}
form a complex of interrelated processes which share some common
features. They can be analysed using chiral perturbation theory
\cite{CPT}, and the experimental exploration of the entire complex
provides stringent checks on the theoretical methods. Predictions for
all of them exist in the literature \cite{EPR,CEP,DG1,VMD}, except for the
one that is the subject of this note, $K_L \rightarrow \pi^0 \gamma
\mu^+ \mu^-$. Our goal is to provide information on this rate and the
corresponding decay distributions.

The motivation for studying this decay can best be seen by first
considering the reaction $ K_L \rightarrow \pi^0 \gamma \gamma $. This
process takes place dominantly through loop process with pions in the
loop. The decay distribution is quite distinctive and the rate is
predicted without any free parameters at one-loop
order. Interestingly, the distribution agrees well with experiment,
but the theoretical rate appears too small by more than a factor of
two. In attempting to resolve this, several authors have gone beyond
the straightforward one-loop (order E$^4$) chiral calculation. One in
particular \cite{CEP} has added a series of higher order effects in a
quasi-dispersive framework and had some surprising success at
increasing the rate without modifying the decay distribution
greatly. The physics which determines $K_L \rightarrow \pi^0 \gamma
\gamma $ also drives the reaction which we consider in this paper. By
studying the leptons plus photon modes experimentally, we are able to
achieve independent confirmation of the dynamics that drives this
complex of decay modes.

We also are motivated to report our results for this mode by the
expectation that it may well be measured in the near future
using KTeV at Fermilab \cite{GW}. Muons have
some experimental advantages compared to electrons, so that although the
present decay channel has a smaller branching ratio, the
mode may be more readily
extracted from the data.

There are two separate calculations that we report on. In the first,
we take only the one-loop result within chiral perturbation theory.
This gives not only a parameter-free prediction for the rate but
also predicts the $q^2/m_\pi^2$ variation of the amplitude with the
invariant mass of the two muons, $q^2$. In the second calculation, we add
all the ingredients considered in Ref. \cite{CEP}, namely higher order behavior
in the experimental $K_L \rightarrow 3 \pi$ decay rate and the effects
of vector meson exchanges between the photon vertices. The latter adds
a free parameter to the problem, so that it is not surprising that we can
modify the rate. However, that parameter is determined by the
$ K_L \rightarrow \pi^0 \gamma \gamma$ rate, so for our process it is
considered already known. If this mechanism is correct, that same parameter
that fixes the rate in one calculation must agree with the rates and
energy distributions in all of the related processes.

The resulting formulas are those already
presented in \cite{DG1}, but in the phase space integration we
substitute the mass of the electron with that of the muon.

In the pure chiral calculation to order E$^4$ the diagrams are shown
in Fig. 1. The result we obtain for the branching ratio up is

\begin{equation}
{\rm B}(K_L \rightarrow \pi^0 \gamma \mu^+ \mu^-) = 4.5 \times 10^{-11}.
\end{equation}

\noindent As in Ref. \cite{DG1}, we define

\begin{equation}
s=(k_1+k_2)^2, \qquad
z = {s \over {m^2_K}}, \qquad
y = {{p^{\phantom{l}}_K \cdot (k_1 - k_2)} \over {m^2_K}}.
\end{equation}

\noindent Here $k_1$ and $k_2$ are the momenta of the off-shell and on-shell
photons, respectively, with the off-shell photon materializing into
the muon pair.

\noindent The decay distributions in $z$ and $y$ provide more detailed
information. We present them in Figs. 2 and 3.

At ${\cal O}$(E$^6$) the higher order effects in the $K_L \rightarrow
3 \pi$ vertex are extracted from a quadratic fit to the
amplitude. These then do not add any new uncertainties to the
radiative amplitude. However, vector meson exchange introduces new
parameters which measure the strength of the exchange diagram of Fig.
4. In Ref. \cite{CEP} the result is parametrized by a ``subtraction
constant'' which must be fit to the $K_L \rightarrow \pi^0 \gamma
\gamma$ decay rate. We include the contributions generated by vector
meson exchange following the procedure outlined in \cite{VMD}. In this
case the calculation leads to a total branching ratio of

\begin{equation}
{\rm B}(K_L \rightarrow \pi^0 \gamma \mu^+ \mu^-) = 5.8 \times 10^{-11}.
\end{equation}

\noindent The decay distributions are presented in Figs. 5 and 6.

\noindent For completeness, we also show the double differential
branching ratios to ${\cal O}$(E$^4$) and ${\cal O}$(E$^6$) in Figs. 7
and 8, respectively.

\section{Conclusions}

The muonic rate is significantly smaller than the corresponding
electronic mode that we studied in Ref. \cite{DG1}. This is of course
due to the more limited phase space, as well as the fact that the
photon propagator is further off-shell in the muonic case. However, we
again see that the more complete calculation presented second above
leads to an enhancement over the purely order E$^4$ calculation
presented first. However, the vector meson diagrams enhance the
muonic mode by a significantly smaller factor than either the original
di-photon mode or the electronic channel. Thus the prediction in the
muonic mode is less useful in deciding on the importance of the vector
exchange, although it will be useful in deciding about the overall
consistency of the calculation scheme.

\bigskip\bigskip
\noindent {\Large \bf Acknowledgments}
\bigskip

\noindent This work is supported in part by the US Department of Energy
under Grant DE-FG05-96ER40945.


\vfill\eject

\centerline{\Large{\bf Figures}}
\bigskip\bigskip\bigskip

\begin{figure}[htb]
\centering
\leavevmode
\centerline{
\epsfbox{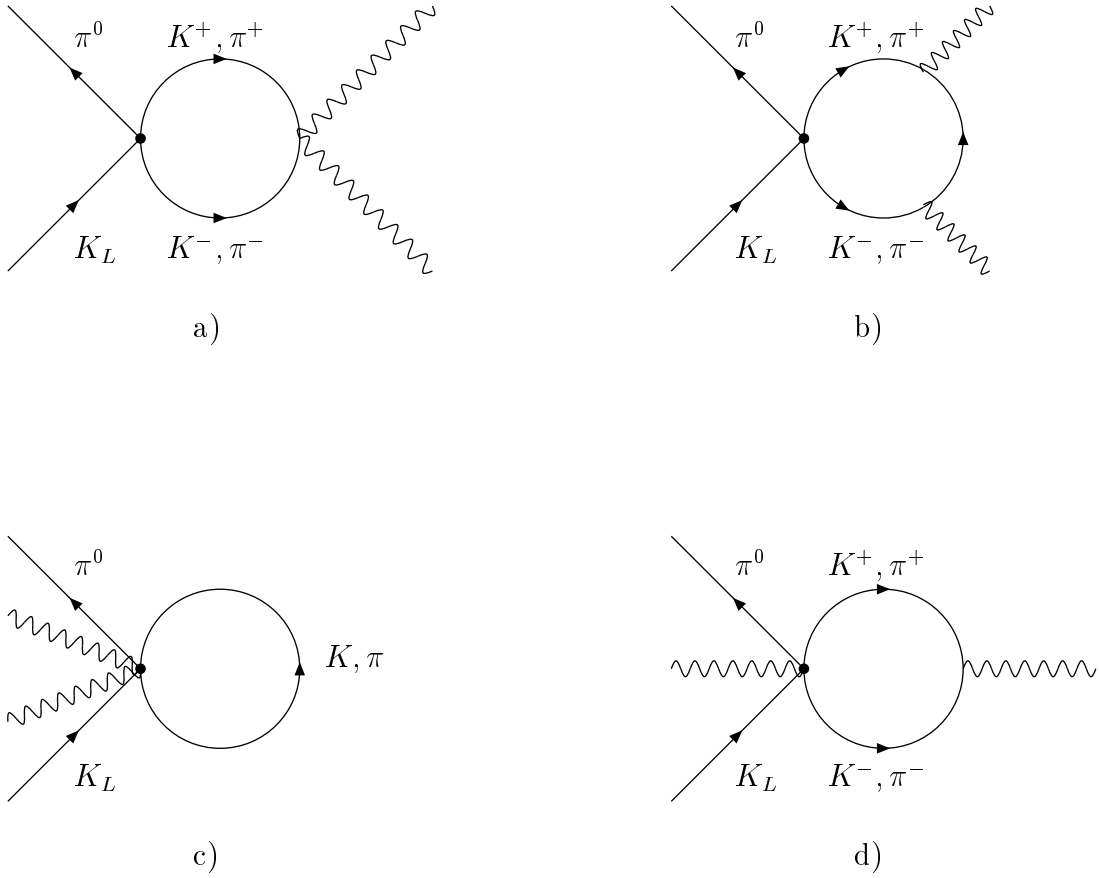}}
\caption{Diagrams relevant to the process $K_L \rightarrow
\pi^0 \gamma \mu^+ \mu^-$ at ${\cal O}(E^4)$ and ${\cal O}(E^6)$.
The muon pair must be attached to one of the photons.}
\end{figure}

\begin{figure}[htb]
\vfill
\centerline{
\begin{minipage}[t]{.47\linewidth}\centering
\mbox{\epsfig{file=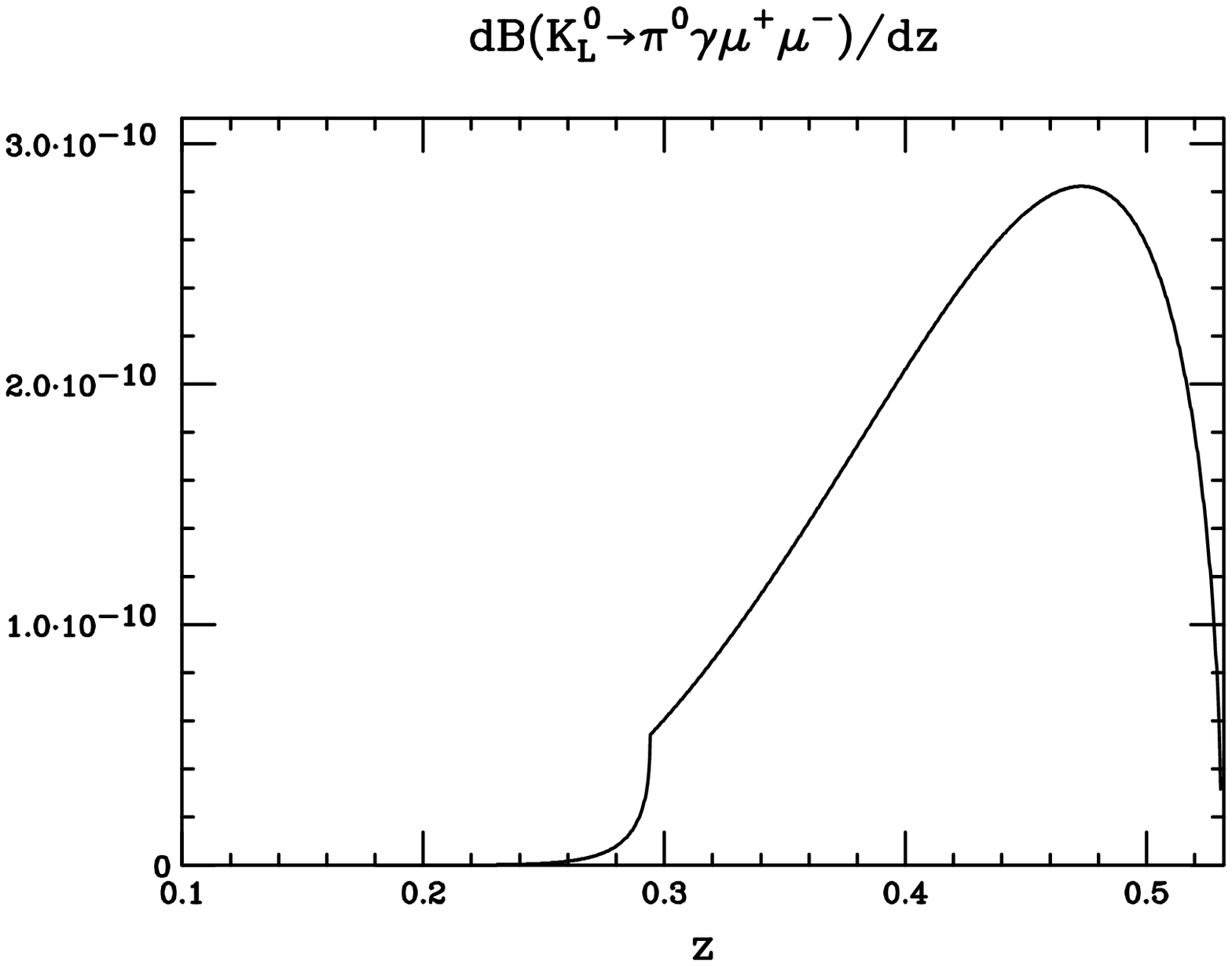,width=8.1cm}}
\caption{The differential branching ratio $dB(K_L \rightarrow
\pi^0 \gamma \mu^+ \mu^-)/dz$ to order E$^4$ is plotted
vs. z.}
\end{minipage}
\hspace{1.0cm}
\begin{minipage}[t]{.47\linewidth}\centering
\mbox{\epsfig{file=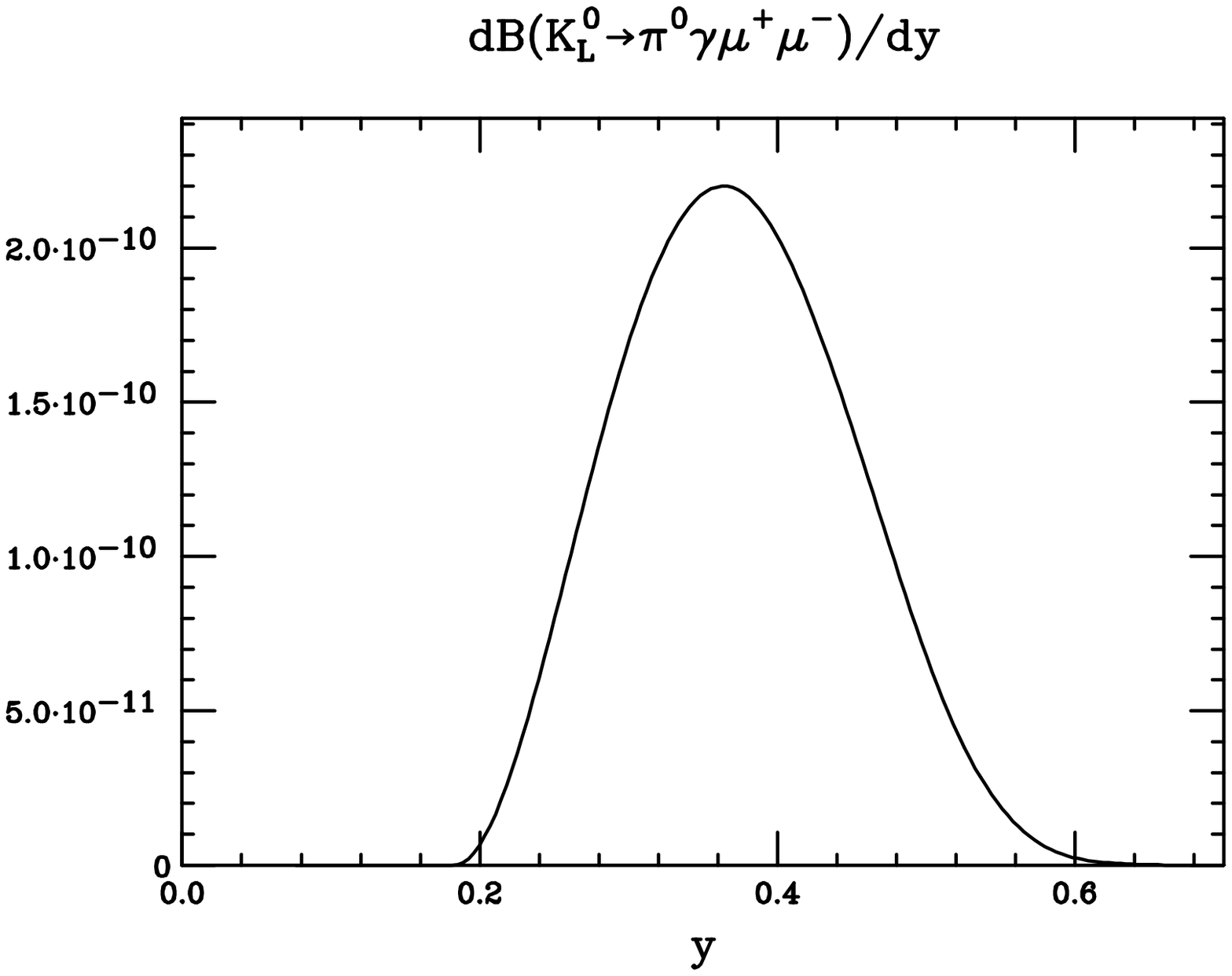,width=8.1cm}}
\caption{The differential branching ratio
$dB(K_L \rightarrow \pi^0 \gamma \mu^+ \mu^-)/dy$
to order E$^4$ is plotted vs. y.}
\end{minipage}}
\end{figure}

\begin{figure}[t]
\centering
\leavevmode
\centerline{
\epsfbox{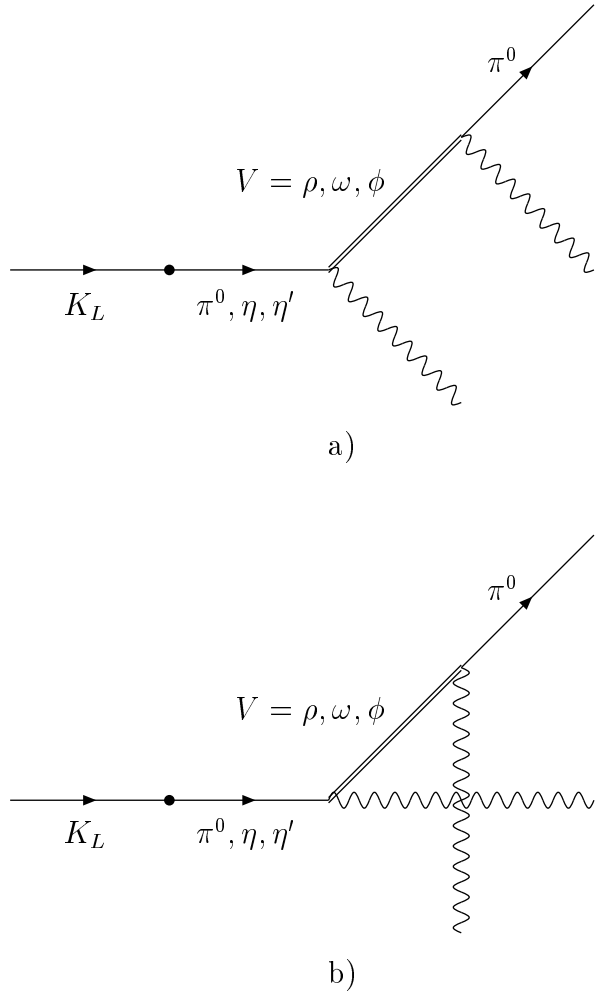}}
\caption{Vector meson exchange diagrams contributing to
$K_L \rightarrow \pi^0 \gamma \mu^+ \mu^-$.}
\end{figure}

\begin{figure}[htb]
\vfill
\centerline{
\begin{minipage}[t]{.47\linewidth}\centering
\mbox{\epsfig{file=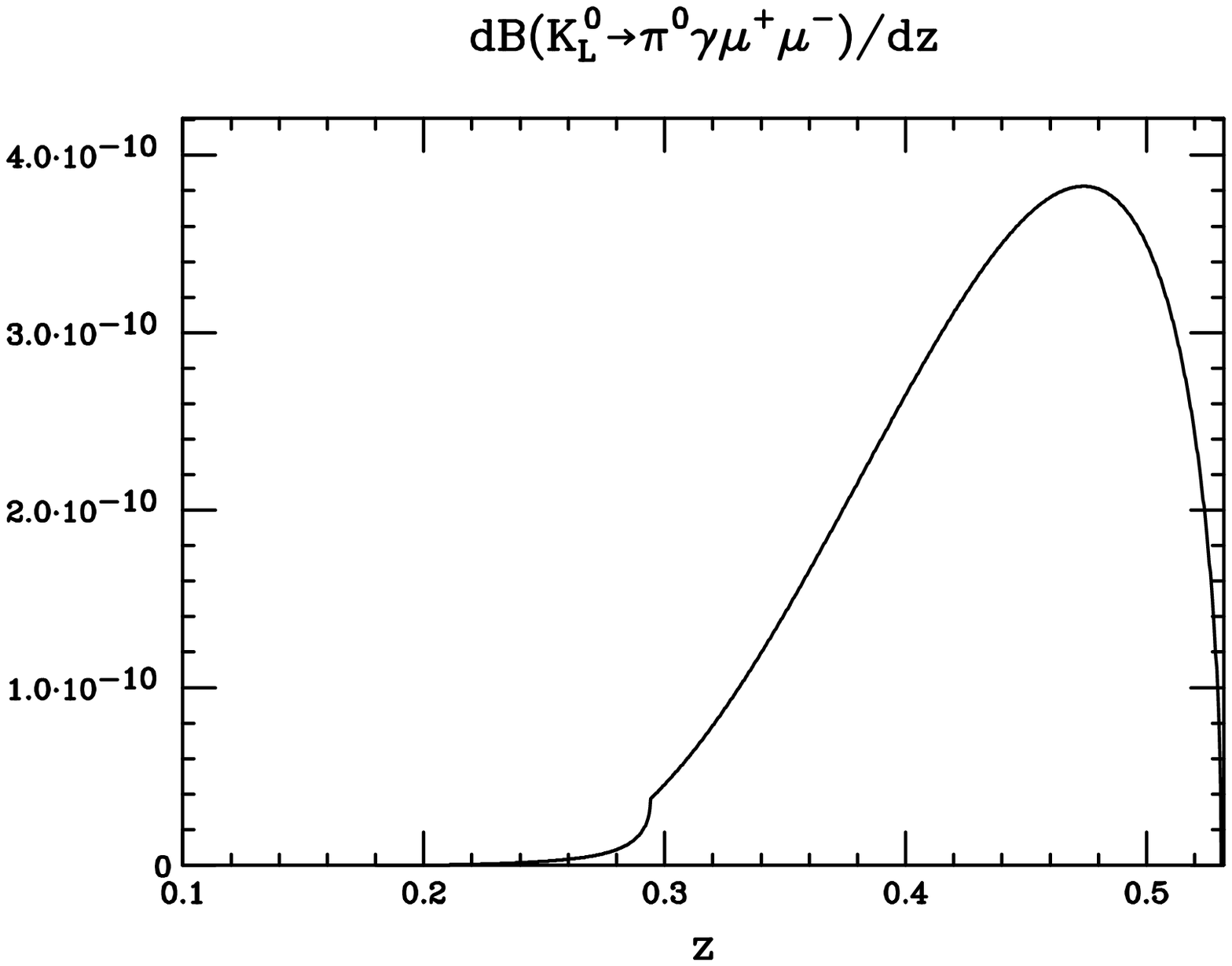,width=8.1cm}}
\caption{The differential branching ratio
$dB(K_L \rightarrow
\pi^0 \gamma \mu^+ \mu^-)/dz$ to order E$^6$ is plotted vs. z.}
\end{minipage}
\hspace{1.0cm}
\begin{minipage}[t]{.47\linewidth}\centering
\mbox{\epsfig{file=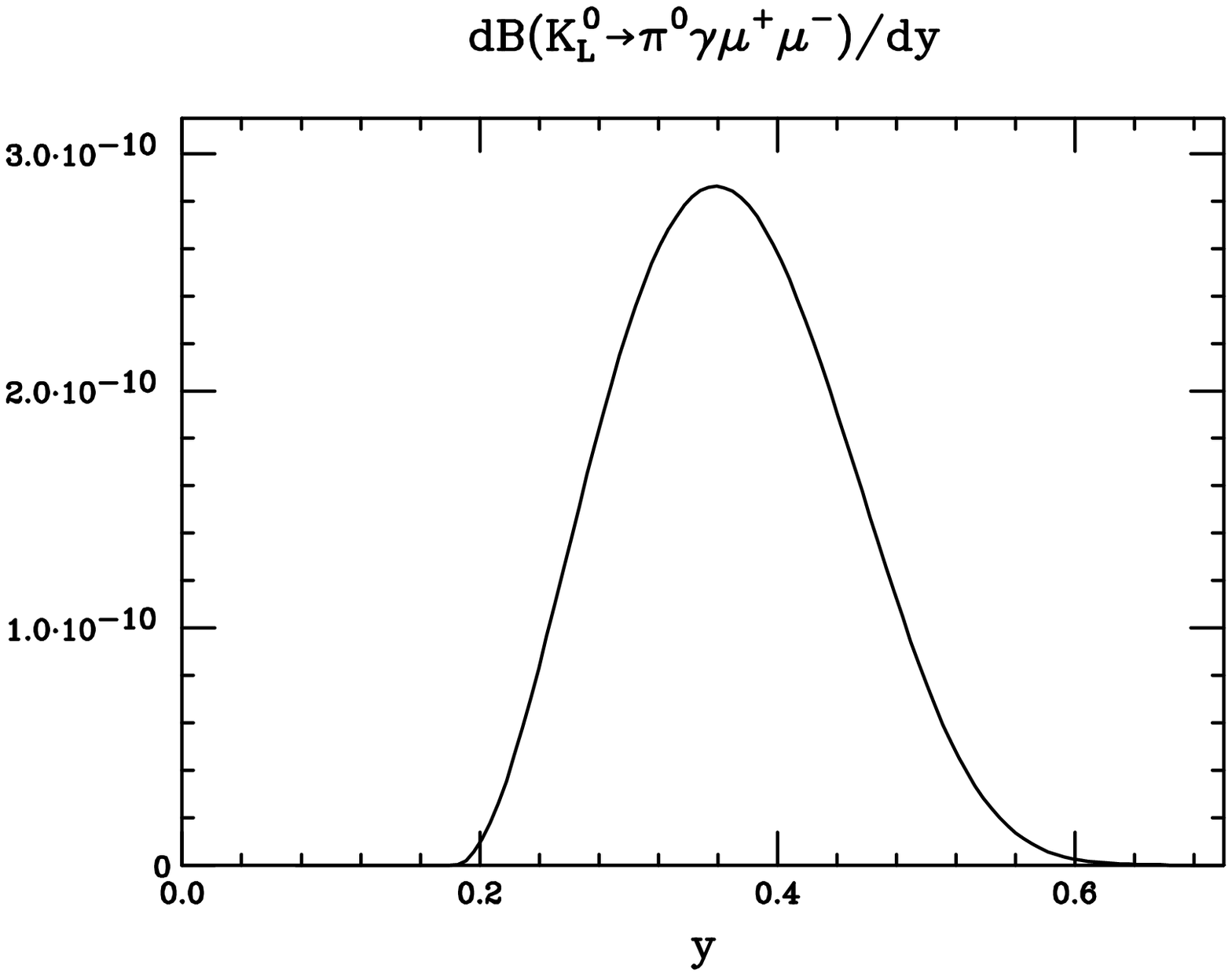,width=8.1cm}}
\caption{The differential branching ratio
$dB(K_L \rightarrow
\pi^0 \gamma \mu^+ \mu^-)/dy$ to order E$^6$ is plotted vs. y.}
\end{minipage}}
\end{figure}

\begin{figure}[htb]
\vfill
\centerline{
\begin{minipage}[t]{.47\linewidth}\centering
\mbox{\epsfig{file=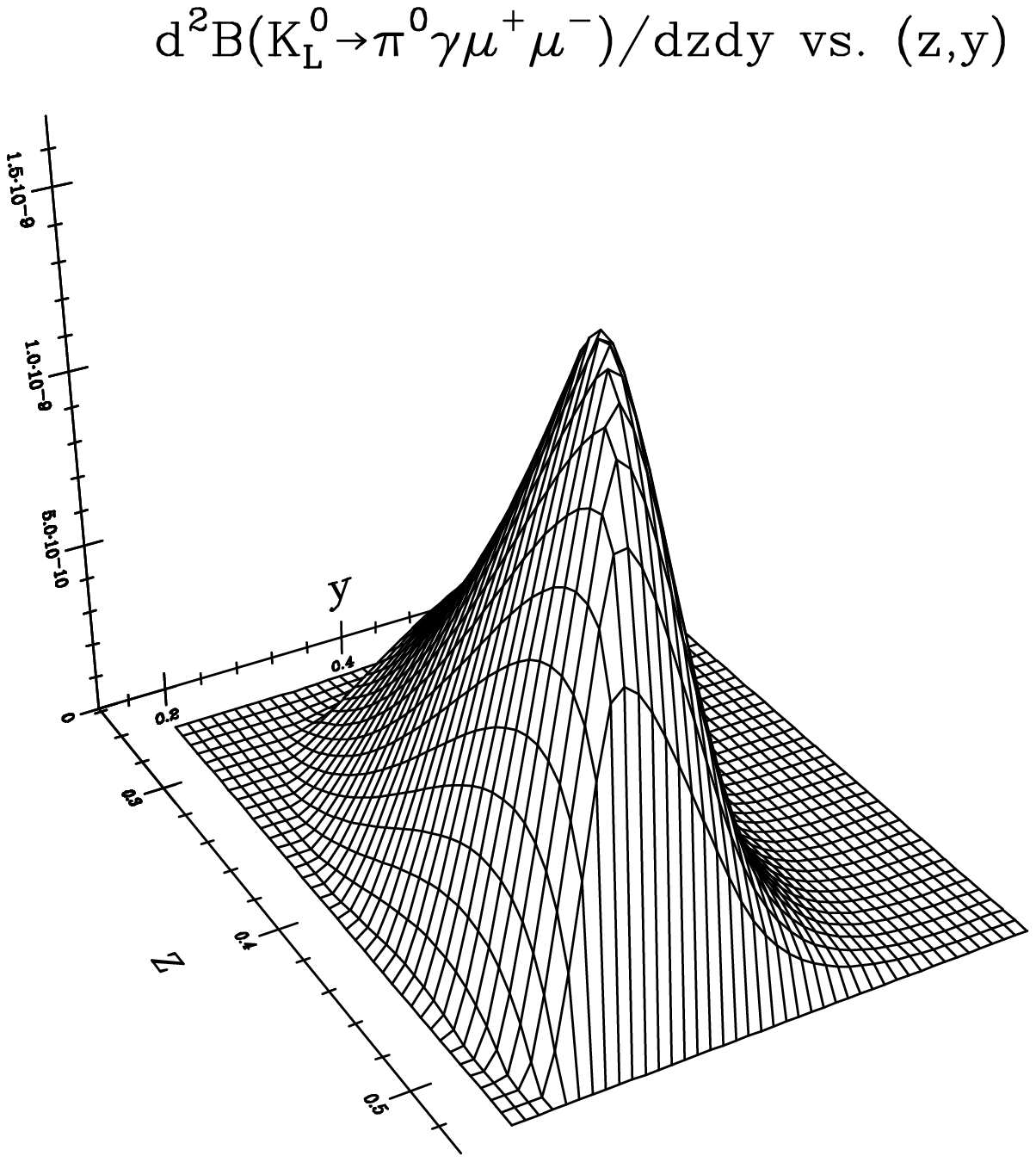,width=8.4cm}}
\caption{The double differential branching ratio
$d^2B(K_L \rightarrow
\pi^0 \gamma \mu^+ \mu^-)/dz dy$ to order E$^4$
is plotted vs. (z,y).}
\end{minipage}
\hspace{0.4cm}
\begin{minipage}[t]{.47\linewidth}\centering
\mbox{\epsfig{file=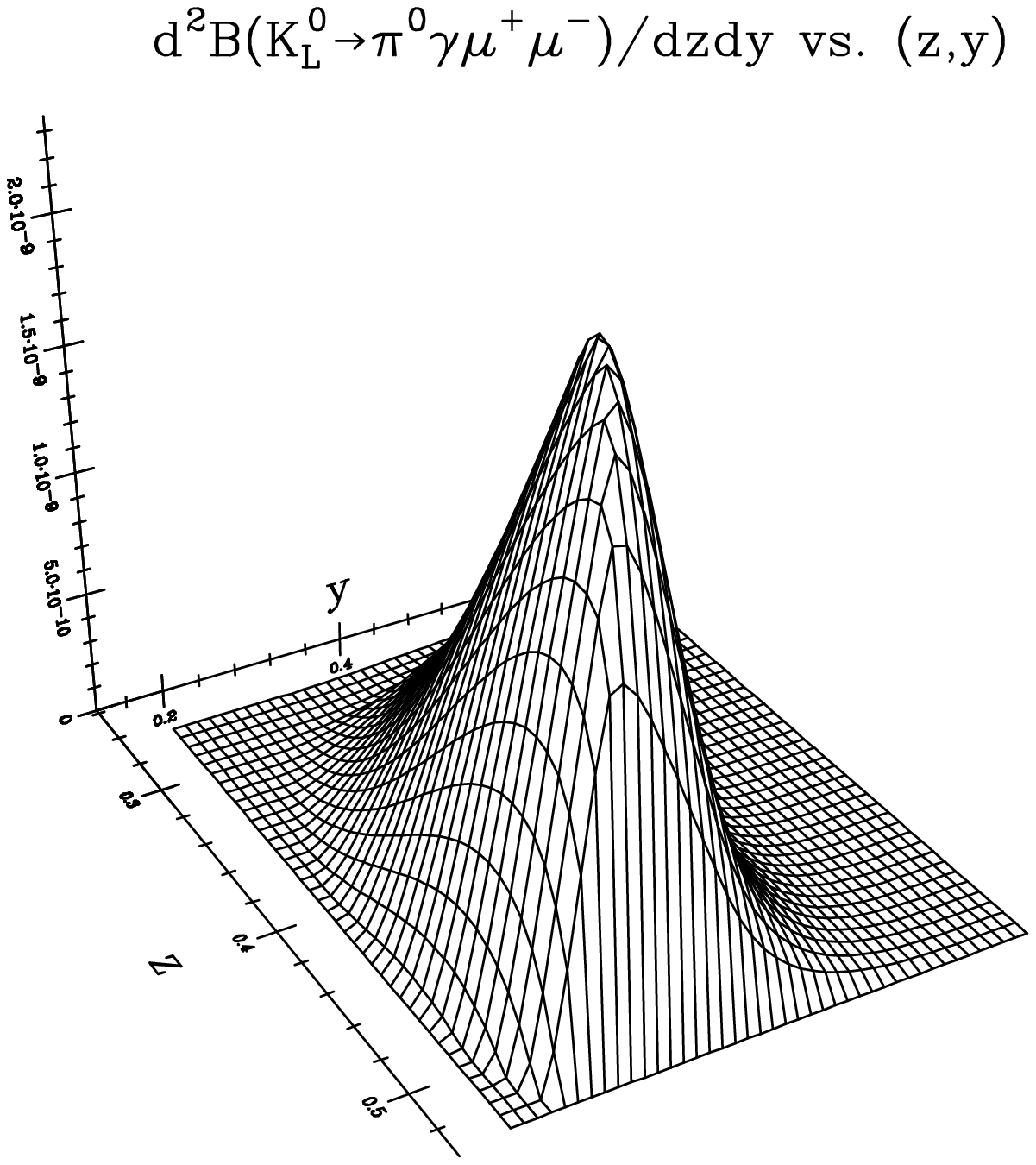,width=8.4cm}}
\caption{The double differential branching ratio
$d^2B(K_L \rightarrow
\pi^0 \gamma \mu^+ \mu^-)/dz dy$ to order E$^6$
is plotted vs. (z,y).}
\end{minipage}}
\end{figure}

\begin{thebibliography}{17}

\bibitem {CPT}
S.~Weinberg, Physica {\bf 96~A}, 327 (1979);
J.~Gasser and H.~Leutwyler, \ap {158} {142} {1984};
\npb {250} {465} {1985}.

\bibitem {EPR}
J.~F.~Donoghue, B.~R.~Holstein and G.~Valencia, \prd {35} {2769} {1987};
G.~Ecker, A.~Pich and E.~de Rafael, \plb {189} {363}
{1987}; {\bf 237}, {481} (1990); \npb {291} {692} {1987};
\npb {303} {665} {1988};
L.~Cappiello and G.~D'Ambrosio, \nc {99} {155} {1988};
L.~M.~Sehgal, \prd {38} {808} {1988};
P.~Ko and J.~Rosner, \prd {40} {3775} {1989};
L.~Cappiello, G.~D'Ambrosio and M.~Miragliuolo, \plb
{298} {423} {1993};
J.~Kambor and B.~R.~Holstein, \prd {49} {2346} {1994};
J.~F.~Donoghue and F.~Gabbiani, \prd {51} {2187} {1995};
G.~D'Ambrosio and J.~Portol\'es, \npb {492} {417} {1997}.

\bibitem {CEP}
A.~G.~Cohen, G.~Ecker and A.~Pich, \plb {304} {347} {1993}.

\bibitem {DG1}
J.~F.~Donoghue and F.~Gabbiani, \prd {56} {1605} {1997}.

\bibitem {VMD}
T.~Morozumi and H. ~Iwasaki, \ptp {82} {371} {1989};
J.~Flynn and L.~Randall, \plb {216} {221} {1989};
L.~M.~Sehgal, \prd {41} {161} {1990};
P.~Ko, \prd {41} {1531} {1990};
J.~Bijnens, S.~Dawson and G.~Valencia, \prd {44} {3555} {1991};
T.~Hambye, \ijm {7} {135} {1992};
P.~Heiliger and L.~M.~Sehgal, \prd {47} {4920} {1993}.

\bibitem {GW} G.~Graham and J.~Whitmore, private communications.

\end{thebibliography}
\end{document}